\documentclass{iau}
\usepackage{graphicx}

\title[JD 11.~~On the Variation of Gas Depletion Time] 
{On the Variation of Gas Depletion Time}

\author[D. Utomo, L. Blitz, A. Bolatto, T. Wong, A. Leroy \& S. Vogel] 
{D. Utomo$^1$, L. Blitz$^1$, A. Bolatto$^2$, T. Wong$^3$, A. Leroy$^4$, S. Vogel$^2$ \and the EDGE-CALIFA Collaboration}

\affiliation{$^1$Department of Astronomy, University of California, Berkeley, CA 94720 
\\[\affilskip]
$^2$Department of Astronomy, University of Maryland, College Park, MD 20742 \\[\affilskip]
$^3$Department of Astronomy, University of Illinois, Urbana, IL 61801 \\[\affilskip]
$^4$Department of Astronomy, The Ohio State University, Columbus, OH 43210}

\pubyear{2015}
\volume{315}  
\jname{From Interstellar Clouds To Star-Forming Galaxies} 
\editors{P. Jablonka, P. Andre \& F. van der Tak, eds.}
\begin{document}

\maketitle

\begin{abstract}
We present an updated status of the EDGE project, which is a survey of 125 local galaxies in the $^{12}$CO($1-0$) and $^{13}$CO($1-0$) lines. We combine the molecular data of the EDGE survey with the stellar and ionized gas maps of the CALIFA survey to give a comprehensive view of the dependence of the star formation efficiency, or equivalently, the molecular gas depletion time, on various local environments, such as the stellar surface density, metallicity, and radius from the galaxy center. This study will provide insight into the parameters that drive the star formation efficiency in galaxies at $z \sim 0$.
\end{abstract}

\firstsection 
\section{Introduction}

The present-day stellar mass of galaxies is an integral of the history of star formation and mergers over the life time of galaxies. Therefore, the rate at which galaxies convert their gas reservoir into stars is one of the important factors that drives galaxy evolution. We have seen in the Milky Way and other nearby galaxies that star formation occurs exclusively in molecular clouds (GMCs, Blitz 1993). In a simplified scheme, these clouds contract under their self-gravity and finally form stars. The contraction time-scale can be estimated to be proportional to the free-fall time, which is a function of gas volume density. Therefore, it is logical to connect the star formation rate (SFR) and the gas volume density (Schmidt 1959). However, the gas volume density is not an observable quantity in extragalactic studies. Instead, we can use the gas surface density. Previous studies showed that SFR surface density correlates with the molecular gas surface density as $\Sigma_{\rm SFR} \propto \Sigma_{\rm mol}^N$ with an exponent of $N \sim 1$ (Bigiel et al. 2008, Leroy et al. 2013).

A linear relation between $\Sigma_{\rm SFR}$ and $\Sigma_{\rm mol}$ implies that the {\it average} molecular gas consumption time-scale or gas depletion time, defined as $\tau_{\rm dep} \equiv \Sigma_{\rm mol}/\Sigma_{\rm SFR}$, is constant, with typical value of $\sim 2-3$ Gyr. The star formation efficiency (SFE) is the inverse of $\tau_{\rm dep}$ and measures the ability of gas to form stars per unit gas mass. Shorter $\tau_{\rm dep}$ means gas forms stars more efficiently. In this paper, we use both terms interchangeably.

The observed scatter in the Kennicutt--Schmidt diagram ($\sim 1$ dex, Kennicutt 1998) means that for a given amount of molecular gas, some galaxies or parts of galaxies (on a $\sim 1$ kpc scale) are more efficient in forming stars than the others by a factor of $\sim 10$. The conditions or parameters that drive the variations of $\tau_{\rm dep}$ are the main issue that is addressed in this paper. In this respect, we need more than just the tracers of SFR (H$\alpha$+IR or UV+IR flux) and molecular gas density (CO flux). Therefore, in $\S 2$ we briefly describe the CALIFA and EDGE surveys that gather other physical properties of galaxies, such as stellar surface densities and metallicities. Then, we analyze if any of these parameters correlate with the SFE or $\tau_{\rm dep}$ in $\S 3$.

\section{The CALIFA and EDGE Surveys}

The Calar Alto Legacy Integral Field Area (CALIFA) is an optical Integral Field Unit (IFU) survey of $\sim 600$ local galaxies at $0.005 < z < 0.03$. 
The samples are selected from the SDSS database based on their diameter in $r$-band ($45'' < D_{25} < 80''$), so that they fit well within the IFU field-of-view. A detailed description of the survey is presented by Sanchez et al. (2012). The main data products of CALIFA survey are various properties and kinematic maps of the stellar populations and ionized gas.

We select 176 galaxies from the CALIFA samples, based on their brightness in the WISE $22\mu$m band, to be observed in the ($J=1-0$) $^{12}$CO and $^{13}$CO spectral lines using the CARMA observatory in the E-array configuration ($8''$ resolution). Then, we chose 125 of them based on $^{12}$CO detections to be observed in the D-array configuration to achieve higher angular resolution ($4''.5$ or $\sim 1-2$ kpc) that is closely matched to the CALIFA resolution ($2''$). This Extragalactic Database for Galaxy Evolution (EDGE) survey is the first major, resolved CO survey matched to an IFU survey. The survey descriptions will be presented elsewhere (Bolatto et al. in prep.). The main data products of the EDGE survey are the surface density and kinematic maps of the molecular gas.

We make use of the CALIFA and EDGE data products to investigate if there is any systematic trend between the depletion time and other galaxy physical properties. We derive the SFR surface densities using the H$\alpha$ line, following the prescription of Calzetti et al. (2007). The H$\alpha$ lines have been corrected for dust extinction by using the Balmer decrement method and the Cardelli et al. (1989) extinction curve. The CO integrated luminosities are converted to the molecular gas surface densities by applying a constant CO-to-H$_2$ conversion factor, $\alpha_{\rm CO}$, of 4.3 $M_{\odot}$ (K km s$^{-1}$ pc$^2$)$^{-1}$. Additionally, we can derive the gas-phase metallicity by using the ratio of strong emission lines (e.g., Kewley \& Dopita 2002), and the stellar mass surface density, $\Sigma_*$, by applying the stellar population synthesis technique to each line-of-sight spectrum. 

\section{Preliminary Results}

We divide our studies into three parts: global (galaxy-by-galaxy), local ($\sim 1 - 2$ kpc scales), and radial analyses as a function of distance from the galaxy centers.

\subsection{Global Variations}

\begin{figure}[t]
\begin{center}
 \begin{tabular}{cccc}
  \includegraphics[width=2.2in]{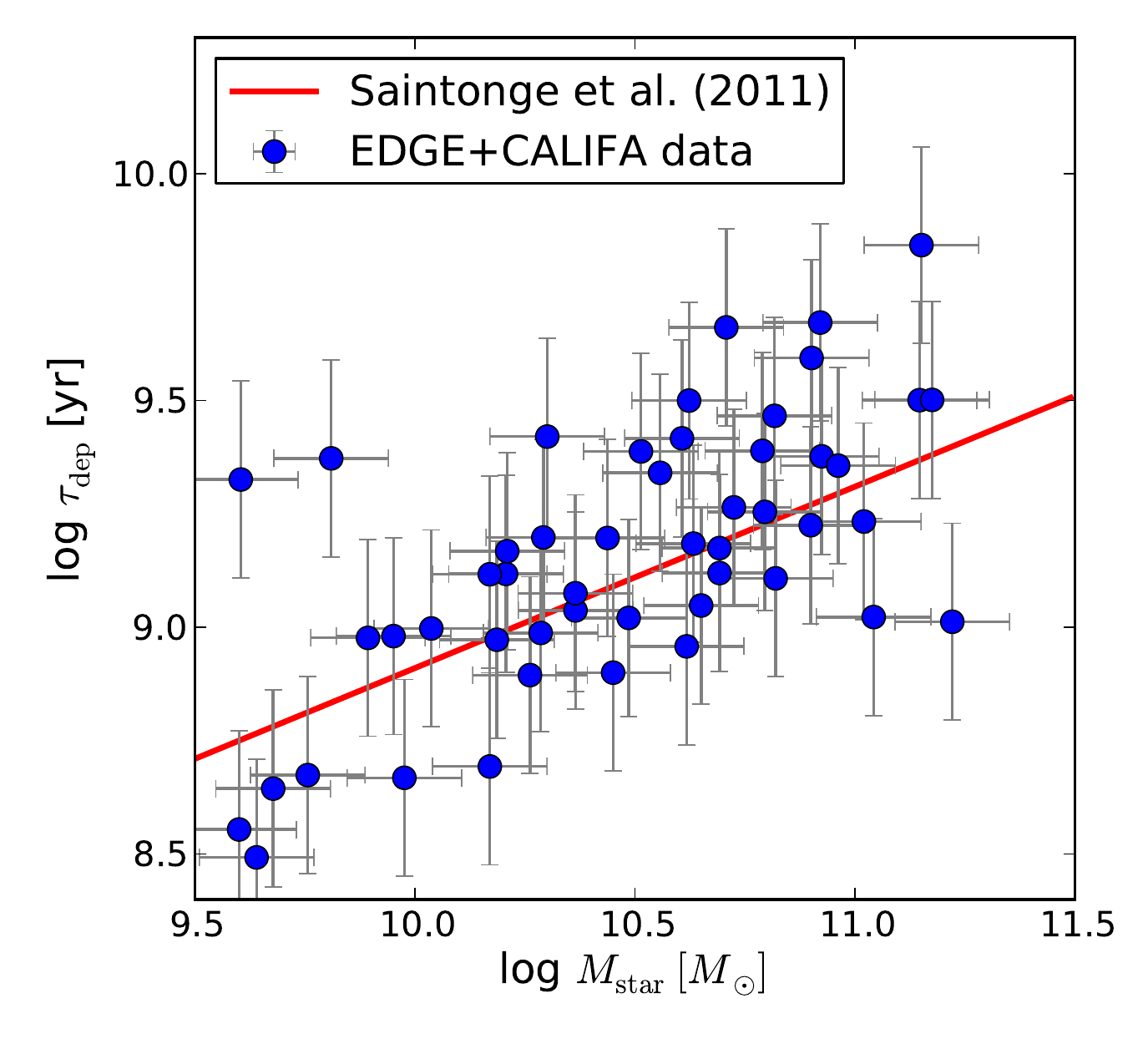} &
  \includegraphics[width=3.05in]{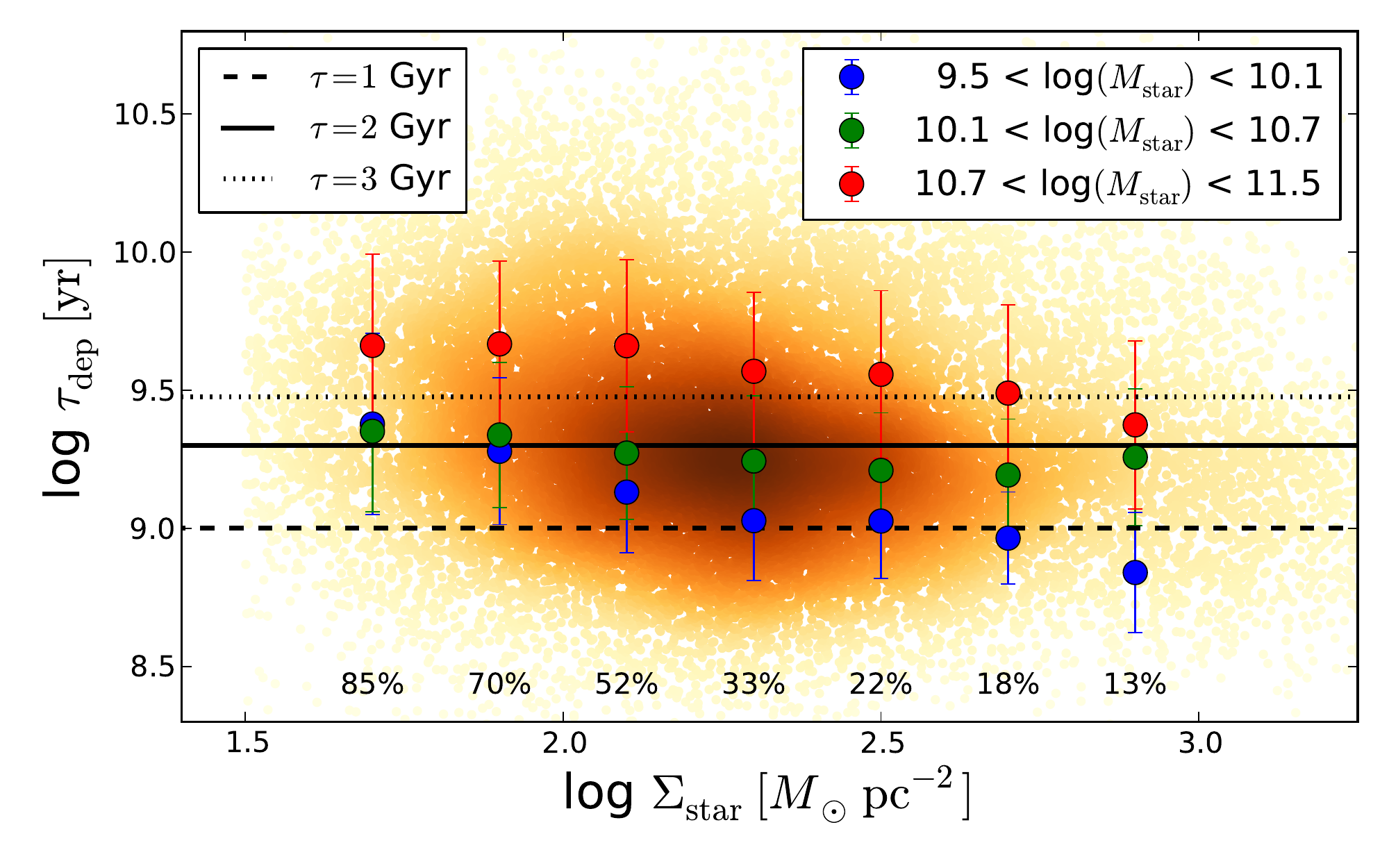} \\
 \end{tabular}
 \caption{Plots of the depletion time vs. stellar mass (left) and stellar surface density (right).}
   \label{fig1}
\end{center}
\end{figure}

Globally, $\tau_{\rm dep}$ increases with the stellar mass, $M_*$ (left panel of Figure\,\ref{fig1}). Here, we compute $\tau_{\rm dep}$ by adding all of the molecular gas mass and dividing it by the sum of all star formation within the galaxy. More massive galaxies tend to have longer $\tau_{\rm dep}$, in agreement with the COLD GASS result (solid red line in Figure\,\ref{fig1}, Saintonge et al. 2011). They argued that the galaxies that show reduced $\tau_{\rm dep}$ are those undergoing minor starbursts due to distant tidal encounters,  variations in the intergalactic medium accretion rate, or secular processes within galactic discs, while morphological quenching and AGN feedback prevent the molecular gas from forming stars in the high mass galaxies.

By using HERACLES samples, Leroy et al. (2013) proposed that the correlation in the $\tau_{\rm dep} - M_*$ diagram can be explained {\it partially} by a variation in $\alpha_{\rm CO}$ with respect to the dust-to-gas ratio, and hence, gas-phase metallicity, $Z$. According to the $M_* - Z$ relation (Tremonti et al. 2004), less massive galaxies have lower $Z$, and therefore, higher $\alpha_{\rm CO}$ (Bolatto et al. 2013). Increasing $\alpha_{\rm CO}$ will shift the low mass galaxies data points upward in the $\tau_{\rm dep} - M_*$ diagram to make a flatter slope. However, the correlation cannot be completely removed, so that a variation in $\alpha_{\rm CO}$ is not the only explanation. A check of how robust our result is with respect to the variation of $\alpha_{\rm CO}$ will be done in a future work (Utomo et al. in prep.).

\subsection{Local Variations}

In order to understand the underlying physics that drive the global trend in $\tau_{\rm dep}$ vs. $M_*$, we make detailed analyses on a scale of $\sim 1-2$ kpc. From many galaxy properties that can be resolved by the IFU observations, we investigate two of them: $\Sigma_*$ and $Z$. These two properties are independent of the $\tau_{\rm dep}$ measurements, but are physically linked to the star formation. Briefly, $\Sigma_*$ is the parameter that dominates the mid-plane pressure, $P_{\rm h}$ (Elmegreen 1989), which correlates to the H$_2$-to-HI mass ratio, $R_{\rm mol}$, as $R_{\rm mol} \propto P_{\rm h}^{0.92}$ (Blitz \& Rosolowsky 2006). 
Higher $\Sigma_*$ raises $R_{\rm mol}$, which in turns produces more sites of star formation (if the molecular gas bound in GMCs). Similarly, higher $Z$ means more shielding of the gas (Wolfire, Hollenbach \& McKee 2010), so that the gas is more likely to produce molecular clouds, which are the sites of star formation. Based on this reasoning, it makes sense to investigate if there is a correlation between $\tau_{\rm dep}$ and $\Sigma_*$ or $Z$.

On the right panel of Figure\,\ref{fig1}, we plot $\tau_{\rm dep}$ vs. $\Sigma_*$. The red, green, and blue dots are the median values of various ranges of $M_*$. At face value, it seems there is a weak correlation so that a higher $\Sigma_*$ has a shorter $\tau_{\rm dep}$. 
This means star formation is more efficient in regions of high $P_{\rm h}$, probably because $P_{\rm h}$ exerts an additional inward force to GMCs so that GMCs are more prone to contract and form stars. Recently, Huang \& Kauffmann (2015) also found this correlation in the HERACLES, ATLAS 3D, and COLD GASS samples. However, they also mentioned the specific SFR (sSFR) as another parameter that drives $\tau_{\rm dep}$, which we do not consider in this paper because it is not an independent parameter with respect to $\tau_{\rm dep}$, i.e. sSFR = SFR/$M_*$.

The scatter in $\tau_{\rm dep}$ itself is driven by different $M_*$ (right panel of Figure\,\ref{fig1}), so that for a given $\Sigma_*$, more massive galaxies have longer $\tau_{\rm dep}$ than less massive galaxies. It is interesting that $M_*$, a global property of galaxies, still leaves its imprint on the local scales. Since the scatter in $\tau_{\rm dep}$ is orthogonal to $\Sigma_*$, the local variations in $\tau_{\rm dep}$ depend on {\it both} $\Sigma_*$ and $M_*$. Note that $\Sigma_*$ here is $\Sigma_*$ on $\sim 1$ kpc scale, not its average value over a whole galaxy, $\bar \Sigma_* \propto M_* \ R_{\rm gal}^{-2}$. Therefore, two galaxies with different $M_*$ can have regions with the same $\Sigma_*$, but located at different radii from the centers due to different absolute scale of their $\Sigma_*$ profile, i.e. it is closer to the center for lower $M_*$ galaxy. 
This leads us to a question of whether $\tau_{\rm dep}$ also varies with radius, which we discuss in $\S 3.3$.

The caveat in the local analyses is that the {\it non-}detection fractions of $\Sigma_{\rm mol}$ (labeled as percentages in Figure\,\ref{fig1}) rise steeply in the lower $\Sigma_*$ regimes. In a future study, we will investigate the effect of non-detections by stacking the CO spectra to increase the sensitivity in the lower $\Sigma_*$ regimes (Utomo et al. in prep.). Another caveat is the variation of $\alpha_{\rm CO}$ due to $Z$. Therefore, we also plot $\tau_{\rm dep}$ vs. $Z$, based on Kewley \& Dopita (2002) prescriptions, in the left panel of Figure\,\ref{fig2}. The median values are shown as blue circles. We do not see a clear correlation, except for the low metallicity regime. This reassures us that our choice of $\alpha_{\rm CO} = 4.3 \ M_\odot$ (K km s$^{-1}$ pc$^2$)$^{-1}$ is reasonable.

\subsection{Radial Variations}

\begin{figure}[t]
\begin{center}
 \begin{tabular}{cccc}
  \includegraphics[width=2.65in]{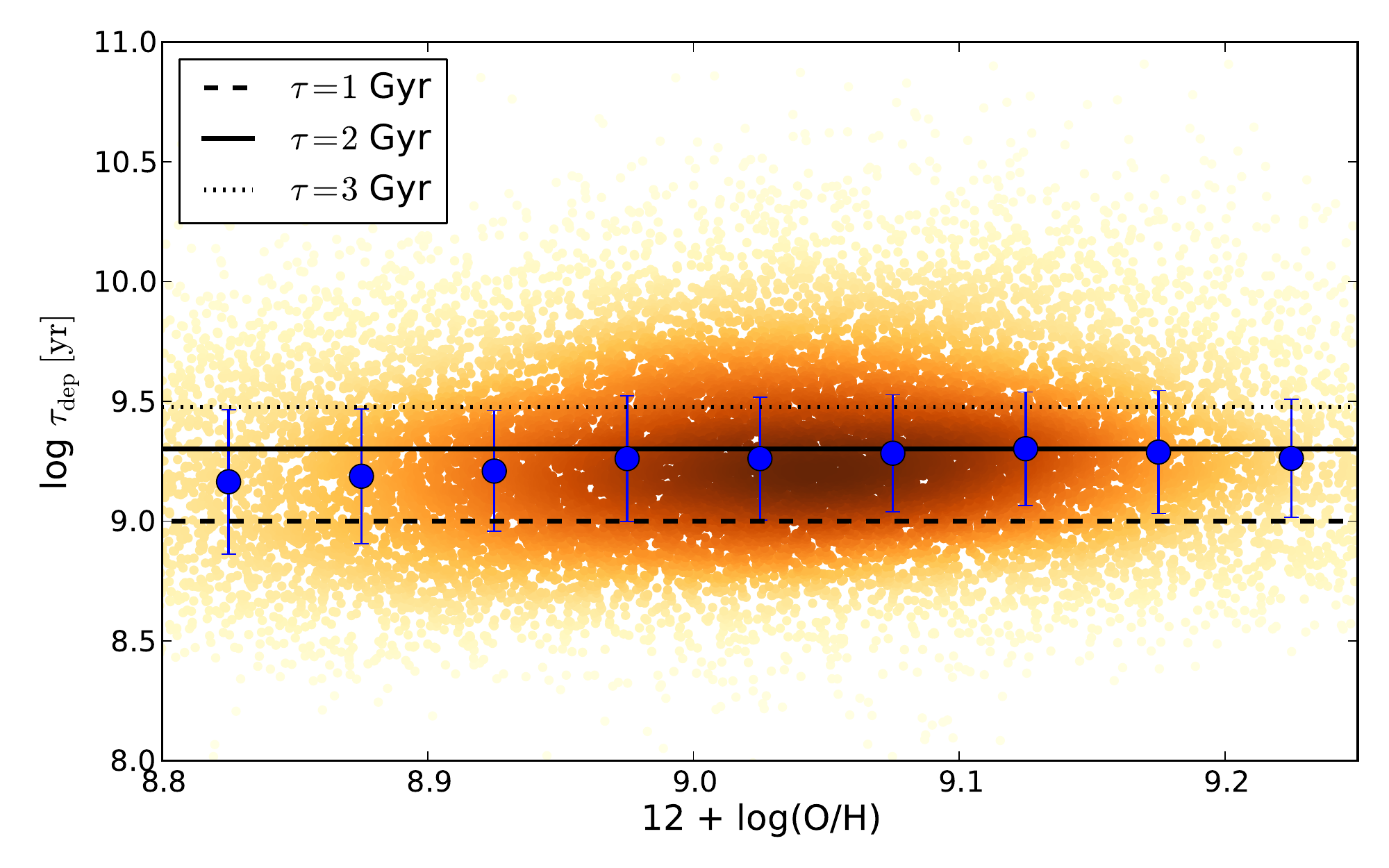} &
  \includegraphics[width=2.65in]{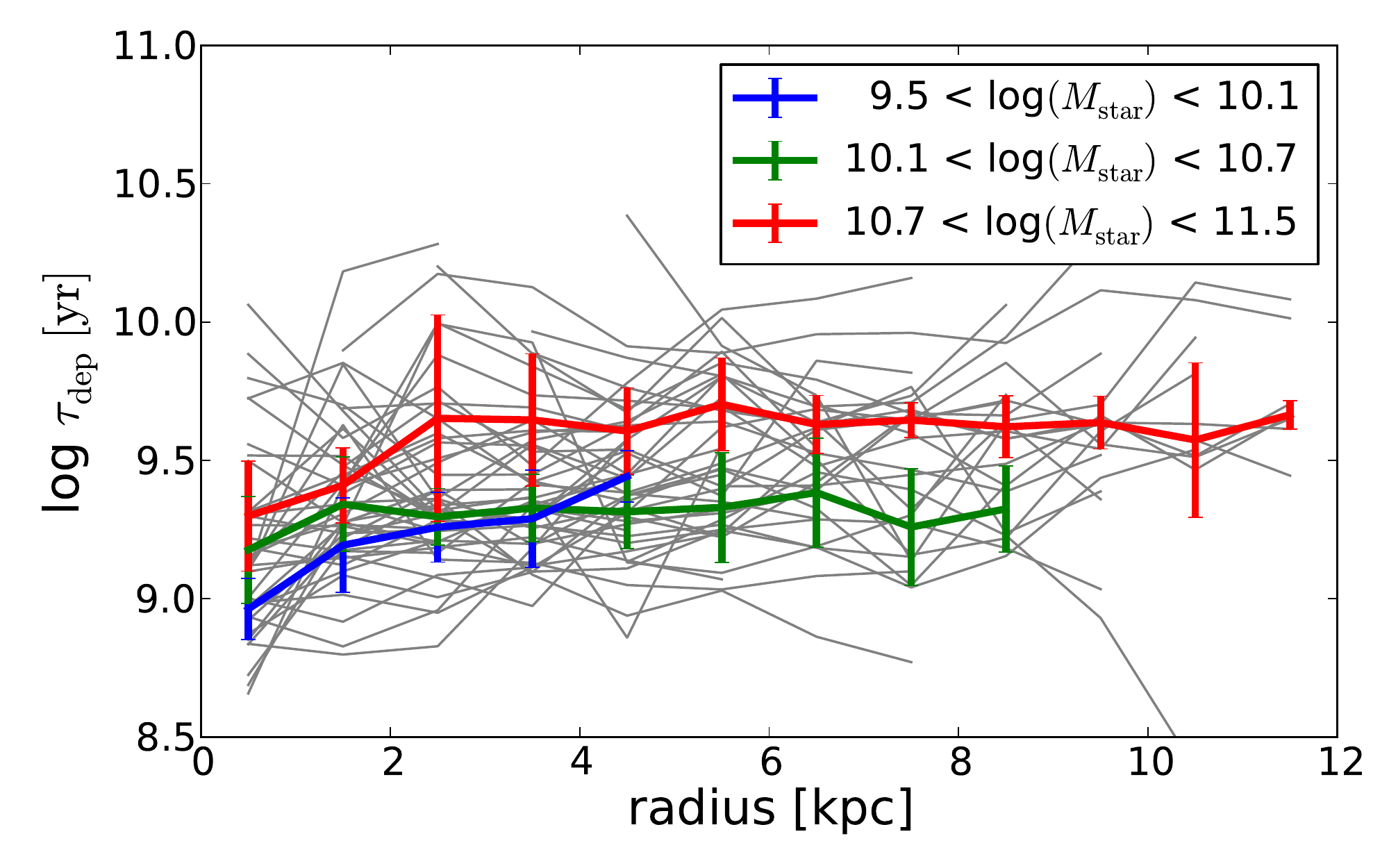} \\
 \end{tabular}
 \caption{{\it Left:} depletion time vs. metallicities. {\it Right:} radial profile of depletion time.}
   \label{fig2}
\end{center}
\end{figure}

Lastly, we investigate the radial profile of $\tau_{\rm dep}$. The radial profiles for individual galaxies are generated by azimuthally averaging $\tau_{\rm dep}$ within a ring. These are shown as the grey lines on the right panel of Figure\,\ref{fig2}. Then, we group the radial profiles into three $M_*$ bins, as in Figure\,\ref{fig1}, and take their median values. We find that the radial profiles of $\tau_{\rm dep}$ are roughly flat for $r \gtrsim 3$ kpc, with an exception for the lowest $M_*$ bin (probably due to the lack of data samples). Interestingly, all three mass bin profiles drop toward the centers at $r \lesssim 3$ kpc, which means the star formation is more efficient in the galaxy centers.

The enhanced SFE at the centers has also been pointed out by \cite[Leroy et al. (2013)]{Leroy13} in the HERACLES samples. They mentioned that this enhanced efficiency coincides with the higher fraction of the $^{12}$CO($2-1$) over $^{12}$CO($1-0$) line intensities, which is an indicator of more excited gas and higher gas temperature at the centers. Furthermore, variations in $\alpha_{\rm CO}$ exaggerate the drop, so this enhancement must be genuine. In the future, we will investigate what causes the drop of $\tau_{\rm dep}$ at the centers, and whether this enhancement of SFE is linked to the morphology of the bulges 
(classical vs. pseudo-bulges and barred vs. unbarred) and the gas free-fall time due to the underlying weight of stars and gas. 



\end{document}